\documentclass[a4paper,11pt]{article}
\usepackage{pos}
\usepackage{braket}
\usepackage{amsmath}
\usepackage{amssymb}
\usepackage{multicol}
\usepackage{slashed}
\usepackage{mathtools}
\usepackage{float}
\usepackage{dsfont}

\let\lpol\l
\usepackage[utf8]{inputenc}\DeclareUnicodeCharacter{2212}{-}
\usepackage[normalem]{ulem}

\title{A lattice QCD determination of the neutron electric dipole moment at the physical point.}
\ShortTitle{A lattice QCD determination of the nEDM at the physical point}

\newcommand{\be}{\begin{equation}}
\newcommand{\ee}{\end{equation}}
\newcommand{\bea}{\begin{eqnarray}}
\newcommand{\eea}{\end{eqnarray}}

\author[a,b]{Constantia Alexandrou}
\author[b,c]{Andreas Athenodorou}
\author[a,b]{Kyriakos Hadjiannakou}
\author*[a,d,e]{Antonino Todaro}

\affiliation[a]{Department of Physics, University of Cyprus, 20537 Nicosia, Cyprus}
\affiliation[b]{Computation-based Science and Technology Research Center, The Cyprus Institute, Cyprus}
\affiliation[c]{Dipartimento di Fisica, Universit\'a di Pisa and INFN, Sezione di Pisa, Largo Pontecorvo 3, 56127 Pisa, Italy}
\affiliation[d]{Faculty of Mathematics und Natural Sciences, University of Wuppertal, 42119 Wuppertal, Germany}
\affiliation[e]{Dipartimento di Fisica and INFN, Universit\'a di Roma ``Tor Vergata'', I-00133 Rome, Italy}

\emailAdd{alexand@ucy.ac.cy}
\emailAdd{a.athenodorou@cyi.ac.cy}
\emailAdd{k.hadjiyiannakou@cyi.ac.cy}
\emailAdd{a.todaro@stimulate-ejd.eu}

\abstract{ Results are presented on the neutron electric dipole moment using an ensemble of $N_f=2+1+1$ twisted mass clover-improved fermions with lattice spacing of $a\simeq 0.08$ fm and physical pion mass ($m_\pi\simeq 139$ MeV). The approach followed in this work is to compute the $CP$-odd electromagnetic form factor $F_3(Q^2\rightarrow0)$ at zero momentum transfer by expanding the action to leading order in $\theta$. This gives rise to correlation functions that involve the topological charge, for which we employ a fermionic definition by means of spectral projectors. We include a comparison between the results using the fermionic and the gluonic definition, where for the latter we employ the gradient flow. We show that using spectral projectors leads to half the statistical uncertainty on the evaluation of $F_3(0)$. Using the fermionic definition, we find a value of $\lvert d_N\rvert = 0.0009(24)\,\theta\, \rm{e}\cdot\rm{fm}$.
}

\FullConference{%
 The 38th International Symposium on Lattice Field Theory, LATTICE2021
  26th-30th July, 2021
  Zoom/Gather@Massachusetts Institute of Technology
}

\begin{document}
\maketitle

\section{Introduction}
We present results on the neutron electric dipole moment (nEDM) induced by a QCD Lagrangian of the form 
\bea
{\cal L}_{\rm QCD} \left( x \right)= \frac{1}{2 g^2} 
{\rm Tr} \left[ F_{\mu \nu} \left( x \right) F_{\mu \nu} \left( x \right) \right] + 
\sum_{f} {\overline \psi}_{f} \left( x \right) (\gamma_{\mu} D_{\mu} + m_f) \psi_{f}\left( x \right)- i \theta q \left( x \right)\,,
\label{eq:QCD_Lagrangian_theta}
\eea
written in Euclidean time. The first two terms are $CP$-conserving while the $\theta$-term is $CP$-violating and thus can give rise to a non-zero nEDM. In the above expression, $\psi_f$ denotes a fermion field of flavor $f$ with bare mass $m_f$, $F_{\mu \nu}$ is the gluon field tensor and $q \left( x \right)$ is the topological charge
density, which in Euclidean space, is defined as 
\bea
  q \left( x \right) = \frac{1}{ 32 \pi^2} \epsilon_{\mu \nu \rho \sigma} 
{\rm Tr} \left[ F_{\mu \nu} \left( x \right) F_{\rho \sigma} \left( x \right) \right]\,.
\label{eq:Topological_Charge_Density}
\eea
Up to date no finite nEDM value has been measured in experiments. The currently best measured upper bound for the nEDM is that given in Ref.~\cite{Abel:2020gbr} as { $\vert \vec{d}_N \vert  < 1.8 \times 10^{-13} e \cdot {\rm fm} \ (90\% \ {\rm CL})$}, measured at the Paul Scherrer Institute (PSI) in Switzerland. From effective field theory calculations~\cite{deVries:2010ah,Mereghetti:2010kp,deVries:2012ab,Guo:2012vf} one derives $\theta \lesssim {\cal O} \left( 10^{-10} -  10^{-11} \right)$ using the experimental bound. However, a direct  determination of the $\theta$-induced nEDM from the QCD Lagrangian of  Eq.~\eqref{eq:QCD_Lagrangian_theta} would require a non-perturbative calculation. The lattice QCD formulation provides an ideal framework to accomplish this task and several attempts to determine the \(\theta\)-induced nEDM are already present in literature~\cite{Guadagnoli:2002nm,Faccioli:2004jz,Shintani:2005xg,Aoki:2008gv,Shintani:2014zra,Shindler:2015aqa,Guo:2015tla,Alexandrou:2015spa,Syritsyn:2018mon,Dragos:2019oxn,Bhattacharya:2021lol}. The method used in our work involves the calculation of the $CP$-odd $F^{\theta}_3(Q^2)$ electromagnetic form factor, that, in the limit of low momentum transfer, gives rise to the following expression for the nEDM~\cite{Pospelov:2005pr}:
\bea
\lvert \vec{d}_N \rvert =  \lim_{Q^2 \to 0} \frac{\vert F^{\theta}_3(Q^2) \vert}{2 m_N}\,,
\label{eq:dN}
\eea
where  $m_N$ denotes the mass of the neutron and $Q^2{=}-q^2$ the four-momentum transfer in Euclidean space ($q{=}p_f-p_i$).
The $\theta$ index on $F_3^\theta$ indicates that the form factor is $\theta$-dependent and it would be zero if $\theta=0$. It is worth noting that the $\theta$-term is imaginary in Euclidian time, thus lattice configurations cannot be generated directly at $\theta\neq 0$. One standard approach to overcome this issue is to expand the expectation value of a general operator ${\braket{\cal O}}_{\theta}$ in terms of $CP-$violating  powers of $\theta$, namely
\be
\braket{\mathcal{O}}_{\theta} = \frac{1}{Z_{\theta}} \int [\text{dU}] [\text{d}\bar{\psi}] [\text{d}\psi] \: \mathcal{O} e^{-S_{QCD}}(1 + i \theta {\cal Q} + {\it O}(\theta^2)) \simeq \braket{\mathcal{O}}_{\theta=0} + i\theta \braket{\mathcal{O}\mathcal{Q}}_{\theta=0} + O(\theta^2)\,,
\label{eq:exp_eitheta}
\ee
where ${\braket{\cal O}}$ is the expectation value computed using the standard $CP-$preserving action with $\theta=0$, and ${\cal Q}$ is the integral over space-time of the topological charge density, which gives the total topological charge.  

As a consequence of this expansion, the two- and three-point functions that enter the computation of $F^{\theta}_3$ are modified by the insertion of the topological charge, that can introduce large statistical fluctuations. 
Therefore, estimating these correlation functions requires a huge amount of statistics, making the study of the nEDM via lattice QCD a notoriously challenging task from the point of view of computational cost. Often, nEDM studies are conducted with ensembles at larger-than-physical pion mass, that are cheaper, and then a chiral continuum extrapolation to the physical point is performed. However, this approach introduces systematics related to the chiral extrapolation, that are difficult to keep under control. In this work, we  perform a first-principles study directly at the physical pion mass and investigate instead the impact of an alternative lattice discretization of the topological charge on the relevant correlation functions. This is motivated by previous studies~\cite{Alexandrou:2017bzk} from which it emerged that different lattice definitions provide compatible results in the continuum limit while, at finite lattice spacing, definitions based on spectral projectors are less affected by cut-off effects. One of the questions in this study is how this impacts the quality of the signal in our determination of the nEDM.

\section{Method}
\label{sec:method}

The method used in this work has been firstly proposed in Ref.~\cite{Shintani:2005xg} and subsequently widely used in Refs.~\cite{Shintani:2014zra,Shindler:2015aqa,Syritsyn:2018mon,Dragos:2019oxn,Bhattacharya:2021lol}. As anticipated in the previous section, it relies on the calculation of the $CP$-odd $F^{\theta}_3(Q^2)$ form factor by treating the $\theta$-parameter as a small perturbation. In the presence of a $CP-$violating term, the matrix element of the electromagnetic current $\braket{ N(p',s') | {\cal J}_{e.m.}^{\mu} | N(p,s)}_{\theta} = \bar{u}^{\theta}_N(p',s') \left[ \Gamma^\mu(q) \right] u^{\theta}_N(p,s)$, can be rewritten in terms of four form factors as follows
\begin{equation}
    {\Gamma}^\mu(q) = F_1(Q^2)\gamma^\mu + \left( F_2(Q^2) + i \gamma_5 F^{\theta}_3(Q^2)\right) \frac{i \sigma^{\mu\nu} q_{\nu}}{2 m_N^\theta}  + F_A(Q^2)\frac{(\slashed{q}q^\mu-q^2\gamma^\mu)\gamma_5}{m^{\theta,2}_N}\,.
    \label{eq:ff_def}
\end{equation}
 The electromagnetic current is given by \({\cal J}_{e.m.}^{\mu} = \sum_f e_f \bar{\psi}_f\gamma^{\mu}\psi_f \), where \(e_f\) is the electric charge of the quark field \(\psi_f\) and \(\bar{u}^{\theta}_N(p',s')\) is the nucleon spinor in the presence of the $\theta$-term. \(F_1(Q^2)\) and \(F_2(Q^2)\) are the Dirac and Pauli electromagnetic form factors respectively, \(F^{\theta}_3(Q^2)\) is the $CP$-odd form factor and \(F_A(Q^2)\) is the anapole form factor, that vanishes for $C$-preserving actions. They are all expressed as functions of the Euclidean four-momentum transfer squared \(Q^2\). On the lattice, the above matrix elements can be extracted from the Euclidean three-point function given by
\begin{equation}
 G_{3pt}^{\mu,(\theta)}(\vec{p}_f,\vec{q},t_f,t_{ins}) \equiv \braket{ J_N(\vec{p}_f,t_f) | {\cal J}_{e.m.}^{\mu}(\vec{q},t_{ins}) | \bar{J}_N(\vec{p}_i,t_i)}_{\theta} \,,
 \label{eq:G3pt_theta}
\end{equation}
where \(J_N(\vec{p}_f,t_f)\), \(\bar{J}_N(\vec{p}_i,t_i)\) are the nucleon interpolating operators that respectively create a nucleon at time \(t_i\) (source) with momentum \(\vec{p}_i\) and annihilate it at time \(t_f\) (sink) and momentum \(\vec{p}_f\). If we insert a complete set of energy and momentum eigenstates and limit ourselves to the ground state contribution,  Eq.\eqref{eq:G3pt_theta} becomes
\begin{align}\nonumber
    G_{3pt}^{\mu,(\theta)}(\vec{p}_f,\vec{q},t_f,t_{ins},t_i) \simeq\; & |Z_{N}^{\theta}|^2 e^{-E_N^f(t_f-t_{ins})}e^{-E_N^i(t_{ins}-t_{i})} \\ 
    & e^{i \alpha^{\theta}_N \gamma_5} \left(\frac{-i \slashed{p}_f + m_N^\theta}{2 E_N^f}\right) \Gamma^\mu(q) \left(\frac{-i \slashed{p}_i + m_N^\theta}{2 E_N^i}\right) e^{i \alpha^{\theta}_N \gamma_5}\,,
    \label{eq:gs_exp2}
\end{align}
with \(E_N^i\equiv E_N(\vec{p}_i) = \sqrt{\vec{p}_i^2 + (m_N^{\theta})^2 }\), \(E_N^f\equiv E_N(\vec{p}_f) = \sqrt{\vec{p}_f^2 + (m_N^{\theta})^2 }\) and $Z^{\theta}_N$ being some unknown normalization coefficient related to the overlap between the interpolating operators and the nucleon state. The appearance of the so called mixing angle \(\alpha^{\theta}_N\) in Eq.~\eqref{eq:gs_exp2} arises due to the $CP$-violation induced by the \(\theta\)-term, that mixes nucleon eigenstates with defined parity. 
As stated in the previous section, we treat \(\theta\) as a small perturbation. For this reason, quantities in the r.h.s of Eq.\eqref{eq:gs_exp2} can be safely replaced by their leading-order terms in $\theta$ expansion, as follows: { $m_N^{\theta} \simeq m_N + \mathcal{O}(\theta^2)\,, Z_N^{\theta} \simeq Z_N + \mathcal{O}(\theta^2) \,$ and $\alpha_N^{\theta} \simeq \alpha_N^{(1)} \theta + \mathcal{O}(\theta^3), F^{\theta}_3(Q^2) \simeq F_3^{(1)}(Q^2)\theta + \mathcal{O}(\theta^3)$},
while higher order contributions can be neglected. Therefore, in the following we will simply refer to the mixing angle and the $CP$-odd form factor as $\alpha_N$ and $F_3(Q^2)$, correspondingly. Using Eq.~\eqref{eq:exp_eitheta} the expectation value in the l.h.s. of Eq.~\eqref{eq:gs_exp2} can be rewritten in terms of \(G_{3pt}^{\mu,(0)}=\braket{ J_N  {\cal J}_{e.m.}^{\mu} \bar{J}_N}_{0}\) and \(G_{3pt,\mathcal{Q}}^{\mu,(0)}=\braket{ J_N  {\cal J}_{e.m.}^{\mu}\, \mathcal{Q} \, \bar{J}_N}_{0}\) that are computed using configurations generated with the standard $CP$-symmetric action. Then, one can relate these two correlation functions to the \(CP\)-even and the \(CP\)-odd parts of the right hand side (r.h.s.) of Eq.~\eqref{eq:gs_exp2}. The form factor can be extracted by choosing an appropriate ratio of three- and two-point functions in order to cancel unknown normalization coefficients. Similarly, one can estimate the nucleon mixing angle \(\alpha_N\) by expanding the relation
\be
G_{2pt}^{(\theta)}(\vec{p}_f,t_f) \equiv \braket{ J_N(\vec{p}_f,t_f) \bar{J}_N(\vec{p}_i,t_i) }_{\theta} = \lvert Z^{\theta}_N \rvert^2 e^{-E_N(t_f-t_i)}\frac{-i\slashed{p}_f+m_N^{\theta}e^{i2\alpha_N^{\theta}\gamma_5}}{2 E_N}\,,
\label{eq:G2pt_theta}
\ee
in powers of \(\theta\). This allows to extract $\alpha_N$ from the two-point functions \(G_{2pt}^{(0)}\) and \(G_{2pt,\mathcal{Q}}^{(0)}\).

\section{Lattice setup}
\label{sec:lattice_setup}

For this work, we employ an ensemble of $N_f=2+1+1$ twisted mass fermions, with 2 degenerate light quarks, up and down, plus the strange and charm quarks as a non-degenerate twisted doublet, at maximal twist. The insertion of a clover term ensure the suppression of cut-off effects, reducing the difference between the mass of the charged and neutral pions~\cite{Alexandrou:2018egz}.
For the gluonic sector, we use the  Iwasaki improved gauge action~\cite{Iwasaki:1985we}, with Symanzik coefficients set to \(c_0=3.648\) and \(c_1= (1-c_0)/8\). All quark masses are tuned close to their physical values. In particular, \(m_\pi=139(1)\) MeV and \(m_N=940(2)\) MeV, the lattice size \(64^3\times 128\) and the  lattice spacing \(a=0.0801(4)\) resulting in  \(m_\pi L=3.62\). We will refer to this ensemble as cB211.72.64. More details about the generation of this ensemble can be found in Ref.~\cite{Alexandrou:2018egz}. 
For the computation of the nucleon two- and three-point functions we employ the standard proton interpolating field $J_N(x) = \epsilon^{abc} \left[ u^{a,T}(x) \mathcal{C}\gamma_5 d^{b}(x) \right]\, u^{c}(x)$, where \(u(x)\) and \(d(x)\) are up and down quark fields in the physical base, and \(\mathcal{C}=i\gamma_2\gamma_4\) is the charge conjugation matrix. Since up and down quarks are degenerate in our formulation, the proton and neutron are degenerate. We use Gaussian smeared quark fields~\cite{Gusken:1989qx,Alexandrou:1992ti}, with \(125\) smearing steps and parameter \(\alpha_G=0.2\), in order to improve the overlap with neutron ground state. Gauge links entering the smearing operator are APE-smeared (\(50\) steps with \(\alpha_{\rm APE}=0.5\) ). For the electromagnetic current \({\cal J}^{\mu}_{e.m.}(x)\)  we use the symmetrized lattice conserved vector current, as defined for instance in Ref.~\cite{Alexandrou:2018sjm}, that does not need renormalization. The reason we symmetrize is to reduce cut-off effects. The explicit expression for the projected two- and three-point are 

\begin{align}
    \label{eq:G2pt_prjcted}
    &G_{2pt}(\Gamma_0,\vec{p}_f,t_f,t_i) \equiv \sum_{\vec{y}} \text{Tr} \left[ \Gamma_0 \braket{ J_N(\vec{y},t_f) \bar{J}_N(\vec{x},t_i)} \right]  e^{-\vec{p}_f(\vec{y}-\vec{x})}\,,  \\
    \label{eq:G2ptQ_prjcted}
    &G_{2pt,\mathcal{Q}}(\gamma_5,\vec{p}_f,t_f,t_i) \equiv \sum_{\vec{y}} \text{Tr} \left[ \frac{\gamma_5}{4} \braket{ J_N(\vec{y},t_f)\, \mathcal{Q} \, \bar{J}_N(\vec{x},t_i)} \right]  e^{-\vec{p}_f(\vec{y}-\vec{x})}\,, \\ 
    \label{eq:G3pt_prjcted}
    &G^{\mu}_{3pt}(\Gamma_k,\vec{q},\vec{p}_f,t_f,t_{ins},t_i) \equiv \sum_{\vec{y},\vec{z}}  \text{Tr} \left[ \Gamma_k \braket{ J_N(\vec{y},t_f) \, {\cal J}_{e.m.}^{\mu}(z,t_{ins})\, \bar{J}_N(\vec{x},t_i)} \right] e^{-\vec{p}_f(\vec{y}-\vec{x})}e^{\vec{q}(\vec{z}-\vec{x})}\,,\\
  \label{eq:G3ptQ_prjcted}
   &G^{\mu}_{3pt,\mathcal{Q}}(\Gamma_k,\vec{q},\vec{p}_f,t_f,t_{ins},t_i) \equiv \sum_{\vec{y},\vec{z}}  \text{Tr} \left[ \Gamma_k \braket{ J_N(\vec{y},t_f)\, {\cal J}_{e.m.}^{\mu}(z,t_{ins}) \, \mathcal{Q} \, \bar{J}_N(\vec{x},t_i)} \right] e^{-\vec{p}_f(\vec{y}-\vec{x})} e^{\vec{q}(\vec{z}-\vec{x})}\,,
\end{align}
where \(\Gamma_0 = \frac{1}{4}(\mathds{1}+\gamma_0)\) and \( \Gamma_k =  i\Gamma_0 \gamma_5\gamma_k\).

At this stage, we consider only connected contributions to the three-point functions. These are computed, using sequential inversions through the sink, at final momentum \(\vec{p}_f=\vec{0}\) and keeping fixed  the sink-source time separation to \(t_{f}-t_{i}=12a\). Larger sink-source separations provide compatible results within errors, and a more careful analysis of the excited states, even if desirable, would require a prohibitively high statistics. For the analysis, we use 750 gauge configurations, separated by 4 trajectories each. We use \(200\) source positions for the computation of the two-point functions for the determination of the mixing angle, corresponding to \(\sim150\)k data, and \(54\) source position (equivalent to \(\sim40\)k data) for the computation of three-point correlators involved in the extraction of $F_3$.

\section{Topological charge}\label{sec:topCharge}

We use two different lattice discretizations for $\mathcal{Q}$. The fist choice has been firstly proposed in Ref.~\cite{DiVecchia:1981aev} and is widely used in literature. It is the discrete counterpart of Eq.~\ref{eq:Topological_Charge_Density} integrated over space-time, where the gluonic field tensor is replaced with a 'clover' term, i.e. a "clover leaf" path \(C_{\mu\nu}\), made by the sum of the plaquettes \(P_{\mu\nu}(x)\) centered in \(x\) and with all the possible orientations in the \(\mu\nu\)-plane, given by
\begin{equation}
    \mathcal{Q}_L = \frac{1}{32\pi^2} \sum_{x} \epsilon_{\mu\nu\rho\sigma} \text{Tr}\left[C_{\mu\nu}(x)C_{\rho\sigma}(x)\right]\,.
    \label{eq:Q_fieldteo_def}
\end{equation}
This operator is even under parity transformations and exhibits \(\mathcal{O}(a^2)\) discretization effects. We use the gradient flow~\cite{Luscher:2010iy} with the standard Wilson action as smoothing action in order to suppress the UV fluctuations of the gauge field that enter \(C_{\mu\nu}\). 
The elementary integration step is \(\epsilon=0.01\) and the topological charge is computed on the smoothed fields at multiples of \(\Delta \tau_{\rm flow}=0.1\). 
The flow time is chosen by studying the dependence of our final quantities on \(\tau_{\rm flow}\) and searching for a plateau region.

The second definition of the topological charge we employed is based on spectral projectors as described in Refs.~\cite{Giusti:2008vb,Luscher:2010ik}. This definition allows one to extract the topological charge from the spectrum of the hermitian Wilson-Dirac operator \(D_W^\dagger D_W\), using the relation
\begin{equation}
    \mathcal{Q} = \frac{Z_S}{Z_P} \sum_{i}^{ \lambda_i < M_0^2 } u_i^{\dagger} \gamma_5 u_i\:,\qquad M_{\rm thr} = Z_P^{-1} M_0\,,
    \label{eq:fQ_ren}
\end{equation}
where \(u_i\) is the eigenvector related to the \(i\)-th eigenvalue \(\lambda_i\), \(Z_P\) and \(Z_S\) are the renormalization constants of the pseudoscalar and scalar densities, respectively, and \(M_0\) is the bare spectral threshold. It bounds the modes that enter into the sum in Eq. \eqref{eq:fQ_ren} by requiring \(\lambda_i<M_0^2\). We use $Z_P=0.462(4)$~\cite{Alexandrou:2019brg} and $Z_S=0.620(4)$ (for details about  our renormalization program see Refs~\cite{Alexandrou:2019ali,Alexandrou:2019brg,Alexandrou:2020sml}). We calculate the lowest 200 eigenvalues of the squared twisted mass Dirac operator using the Implicitly Restarted Lanczos Method (IRLM) where polynomial acceleration is employed. Using the values of $Z_S$ and $Z_P$ quoted above, this corresponds to a threshold \(M_{\rm thr}\) that varies in the range \(0\div65\ \text{MeV}\).  In the rest of the paper, we will refer to the definition of Eq.~\eqref{eq:Q_fieldteo_def} as ``gluonic" or ``field theoretic" definition of the topological charge, while the one defined by Eq.~\eqref{eq:fQ_ren} will be referred to as  the ``fermionic" or ``spectral projectors" definition.

\section{Results}
\label{sec:results}

\begin{figure}[!h]
\begin{minipage}{0.48\linewidth}
    \includegraphics[width=1.\textwidth]{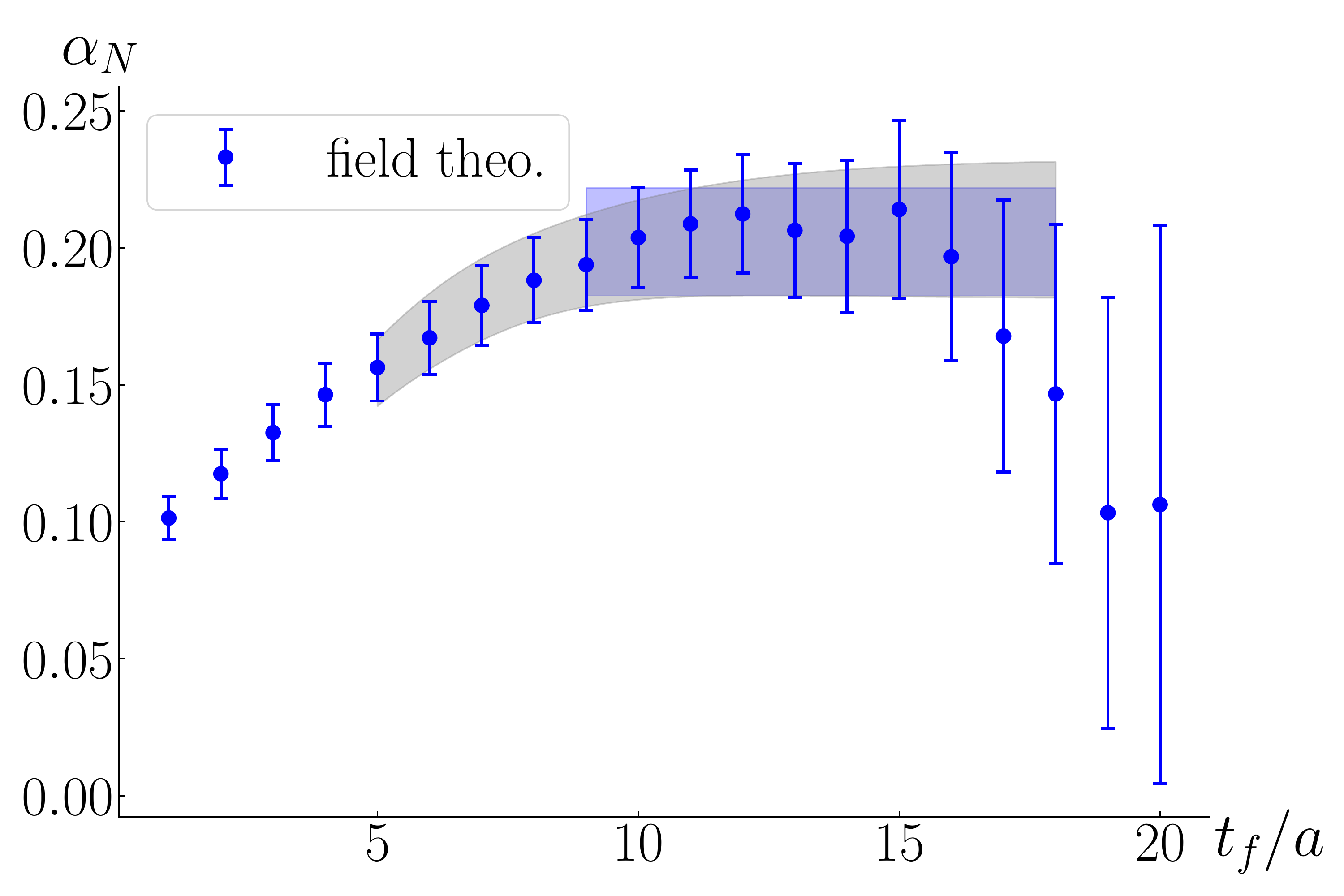}
\end{minipage}%
\hspace*{\fill}
\begin{minipage}{0.48\linewidth}
    \includegraphics[width=1.\textwidth]{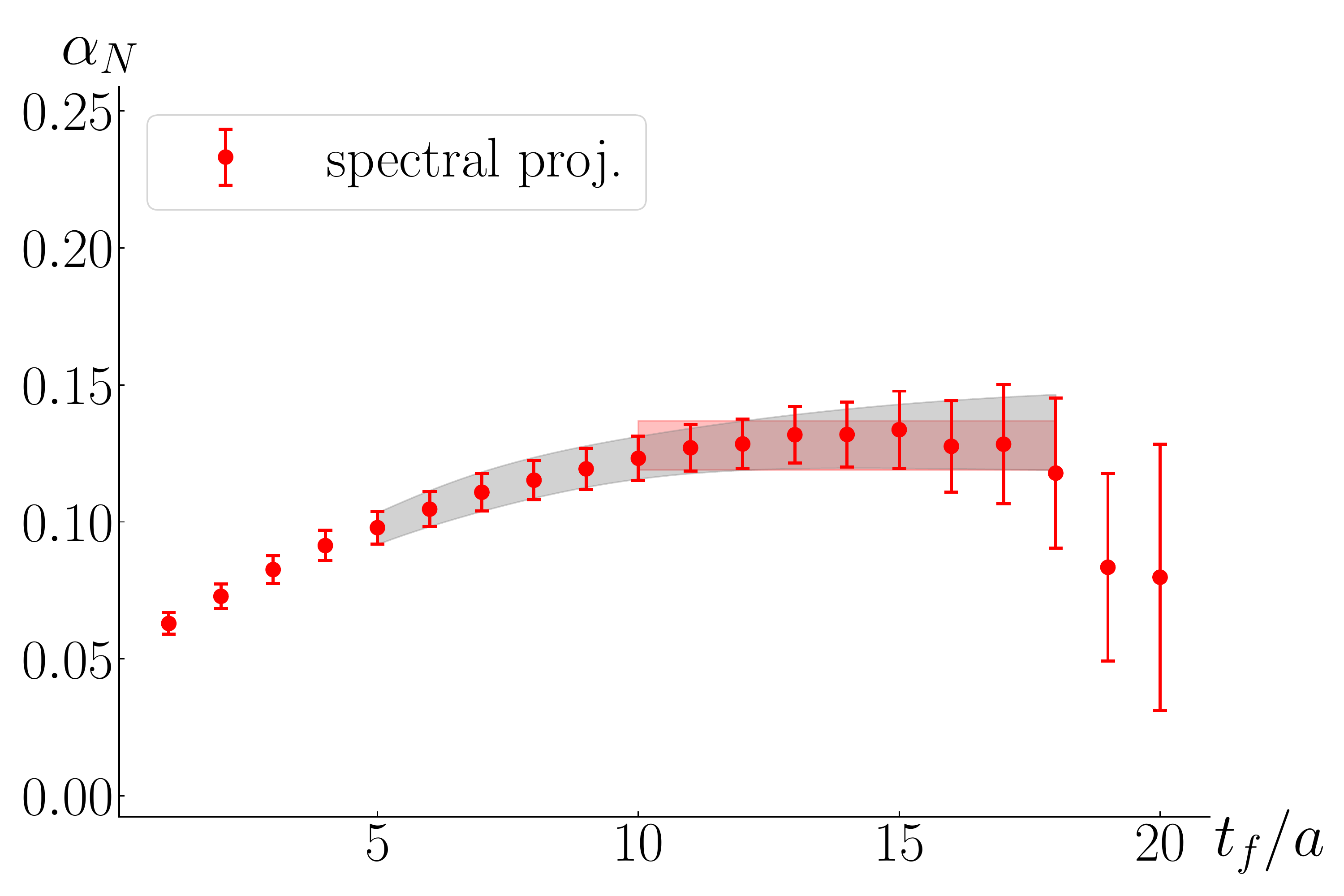}
\end{minipage}
\vspace*{-10pt}
\captionof{figure}{Value of the ratio in Eq.~\eqref{eq:extract_alfa}, as a function of $t_f/a$, using the two definitions of the topological charge, gluonic definition of Eq.~\eqref{eq:Q_fieldteo_def} at \(\tau_{\rm flow}=3.5\) (left) and fermionic definition of Eq.~\eqref{eq:fQ_ren} with \(M_{\rm thr}=64.98\) MeV (right). With the blue (left) and red (right) bands we show the result of a constant fit within the plateau. With the grey band we show the corresponding fits when using for the fit a constant plus an exponential term which takes into account the first excited state.}
\label{fig:alfa_angle_tf}
\end{figure}

For a discussion of the behavior of the topological charge see Ref.~\cite{Alexandrou:2020mds}. The mixing angle \(\alpha_N\) is extracted from the following ratio of two-point functions at zero-momentum
\begin{equation}
    \alpha_N = \lim_{t_f\rightarrow\infty} \frac{G_{2pt,\mathcal{Q}}(\Gamma_5,\vec{0},t_f)}{G_{2pt}(\Gamma_0,\vec{0},t_f)}\,,
    \label{eq:extract_alfa}
\end{equation}
where \(G_{2pt,\mathcal{Q}}\) and \(G_{2pt}\) are defined in Eq.~\eqref{eq:G2ptQ_prjcted} and Eq.~\eqref{eq:G2pt_prjcted}, and $t_i=0$ thus the dependence on \(t_i\) is suppressed. 
The  ratio in Eq. \eqref{eq:extract_alfa} is illustrated in Fig. \ref{fig:alfa_angle_tf}  as a function of \(t_f/a\). We use both the gluonic (left panel) and the fermionic (right panel) definitions for the topological charge.

We seek for a plateau region that we identify in the range \(t_{f}/a \in [9,18]\) and \(t_{f}/a \in [10,18]\) for the  gluonic and fermionic definitions, respectively. Results of constant fitting are represented by the coloured boxes in Fig.~\ref{fig:alfa_angle_tf} and read \(\alpha_N=0.202(20)(4)\) for the gluonic topological charge and \(\alpha_N=0.128(9)(3)\) for the fermionic one, where errors are respectively the statistical and the systematic one. The latter is obtained by varying the initial time slice of the plateau range in the interval \([8,12]\) and the final one in \([17,20]\) and taking the largest difference between the mean values. For both definitions of \(\cal Q\), it is negligible if compared to the statistical uncertainty.
We also tried an exponential Ans\"atz (grey band of Fig.~\ref{fig:alfa_angle_tf}) that provides compatible results, further validating the choice of the plateau region.

\begin{figure}[!h]
\begin{minipage}[c]{0.32\linewidth}
\includegraphics[width=0.88\textwidth]{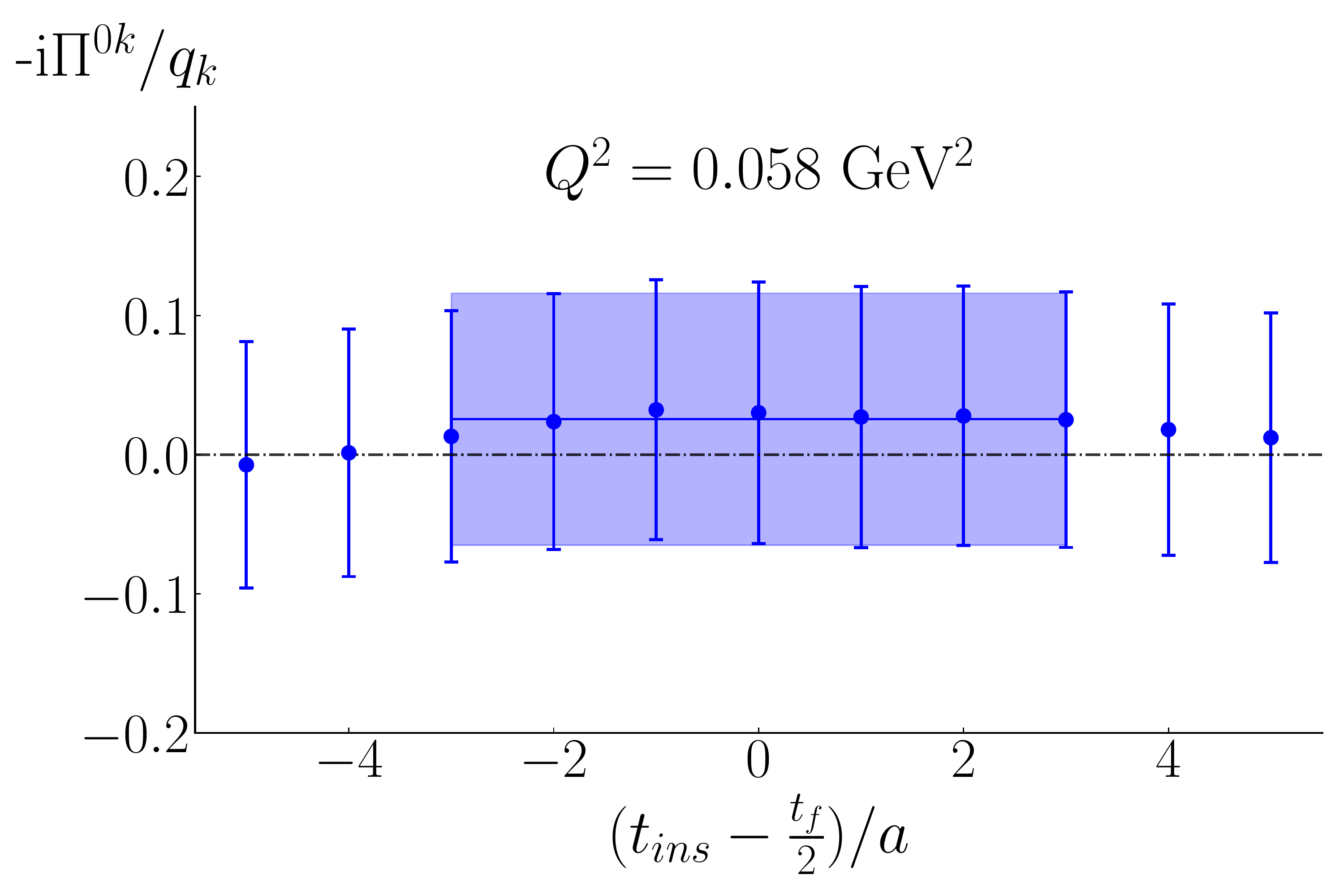}    
\end{minipage}%
\hfill
\begin{minipage}[c]{0.32\linewidth}
\includegraphics[width=0.9\textwidth]{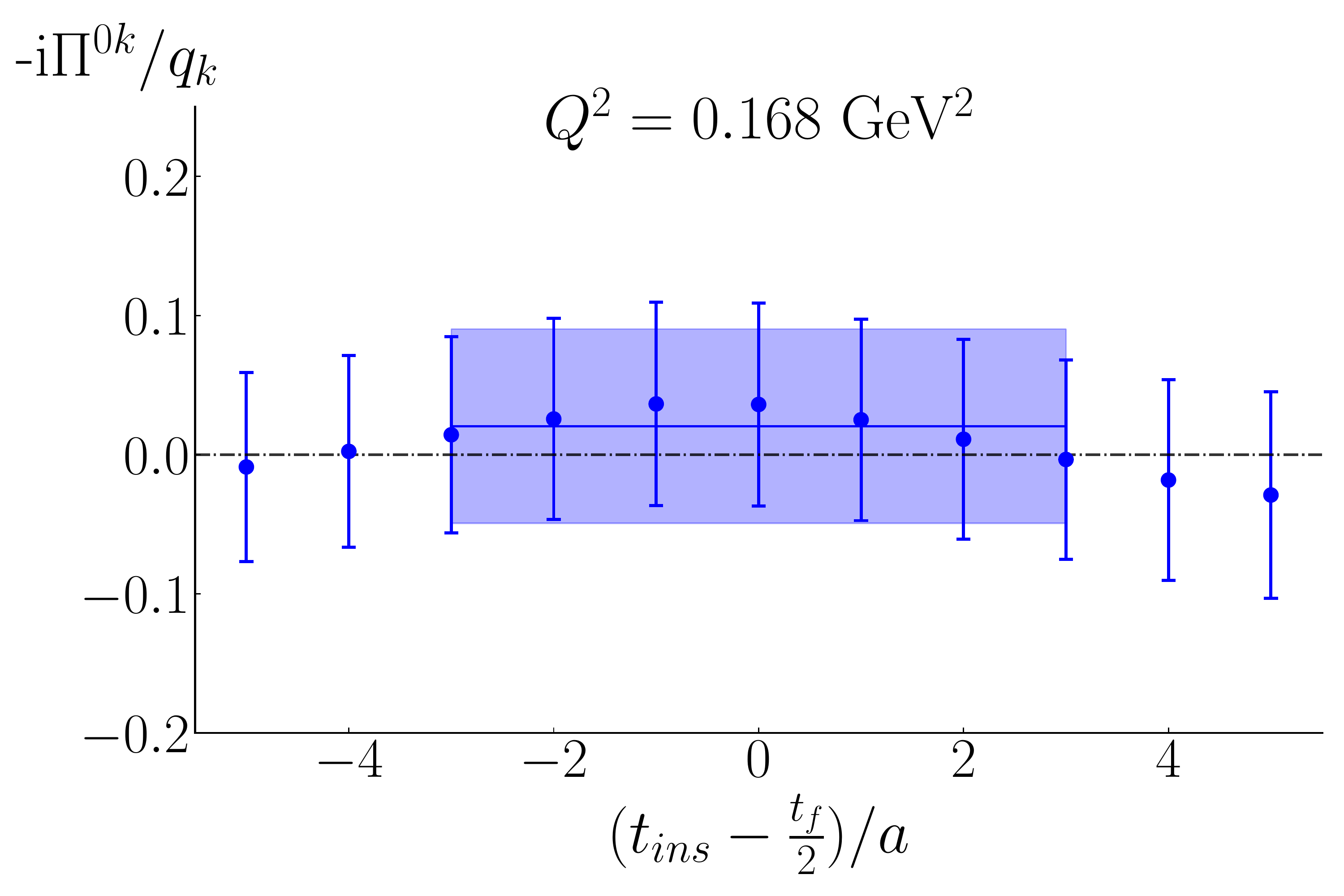}    
\end{minipage}%
\hfill
\begin{minipage}[c]{0.32\linewidth}
\includegraphics[width=0.9\textwidth]{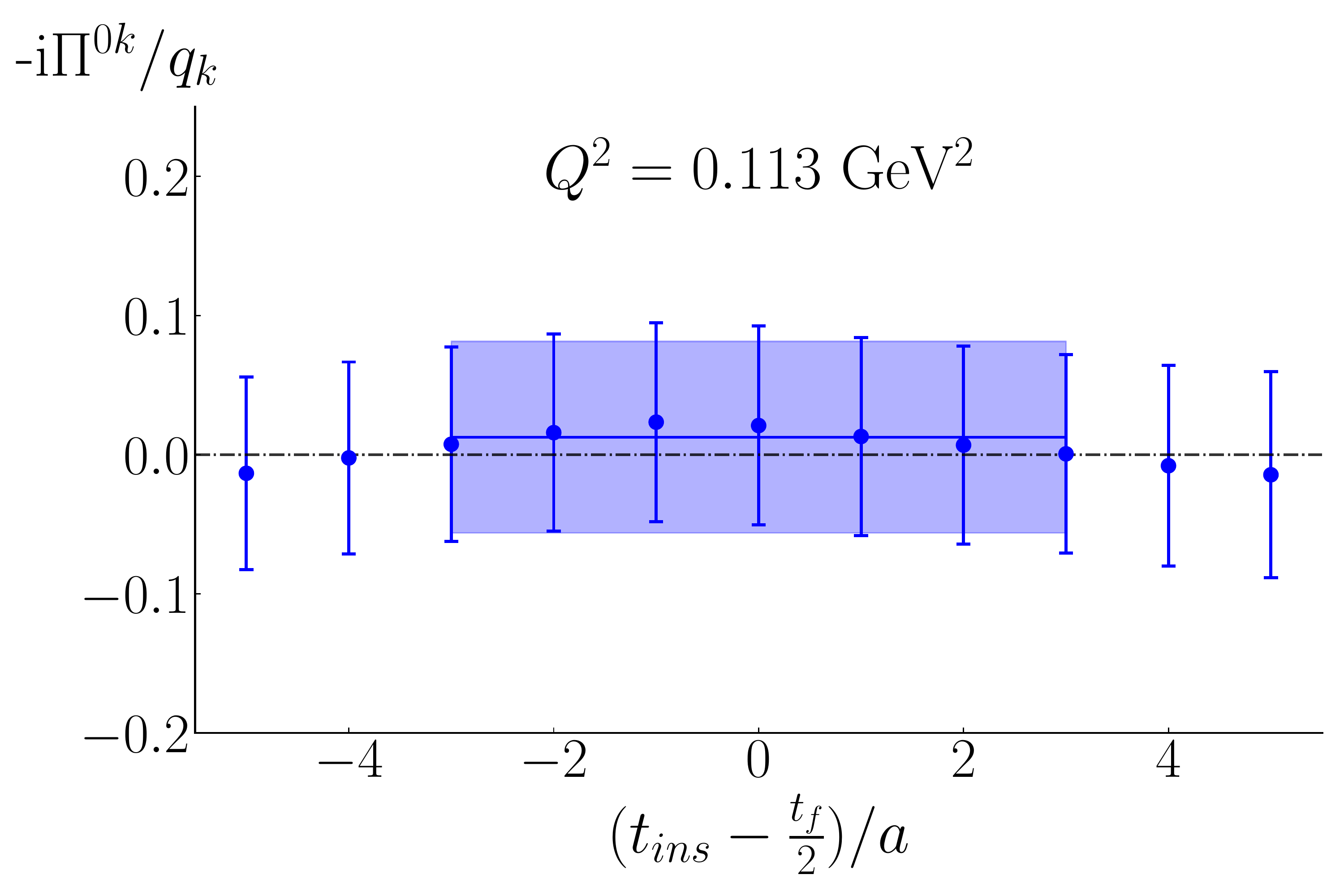}    
\end{minipage}
\begin{minipage}[c]{0.32\linewidth}
\includegraphics[width=0.9\textwidth]{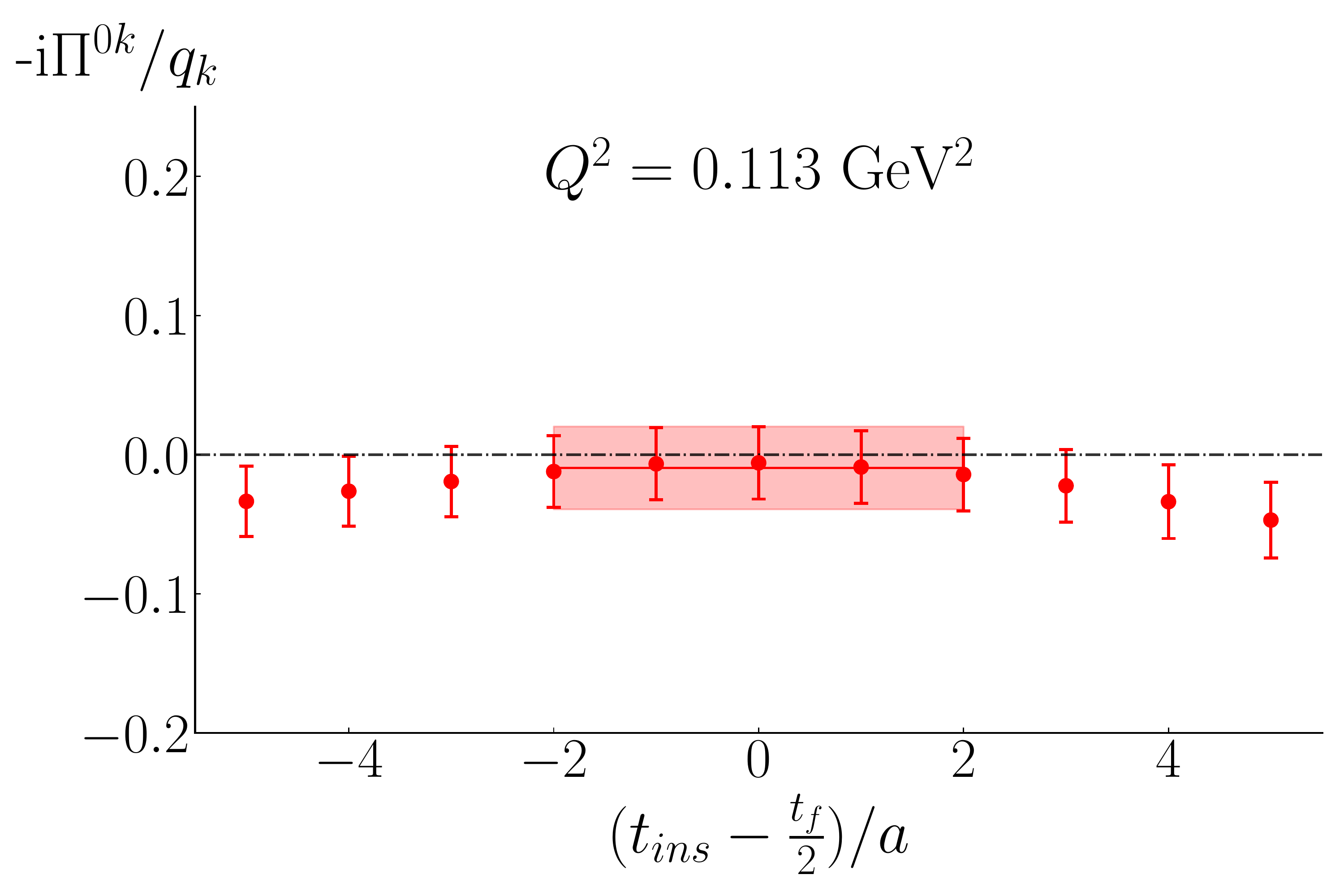}    
\end{minipage}%
\hfill
\begin{minipage}[c]{0.32\linewidth}
\includegraphics[width=0.9\textwidth]{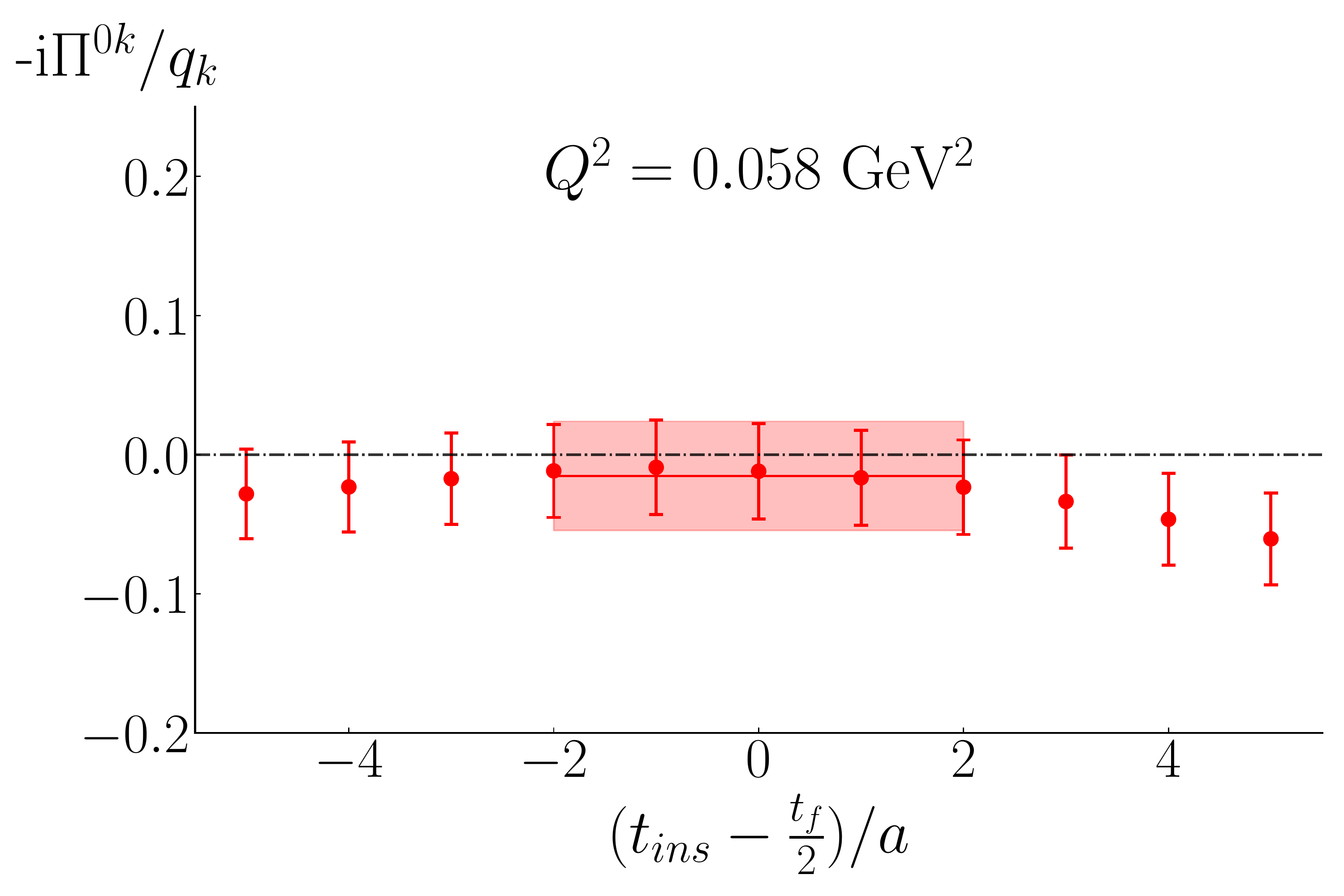}    
\end{minipage}%
\hfill
\begin{minipage}[c]{0.32\linewidth}
\includegraphics[width=0.9\textwidth]{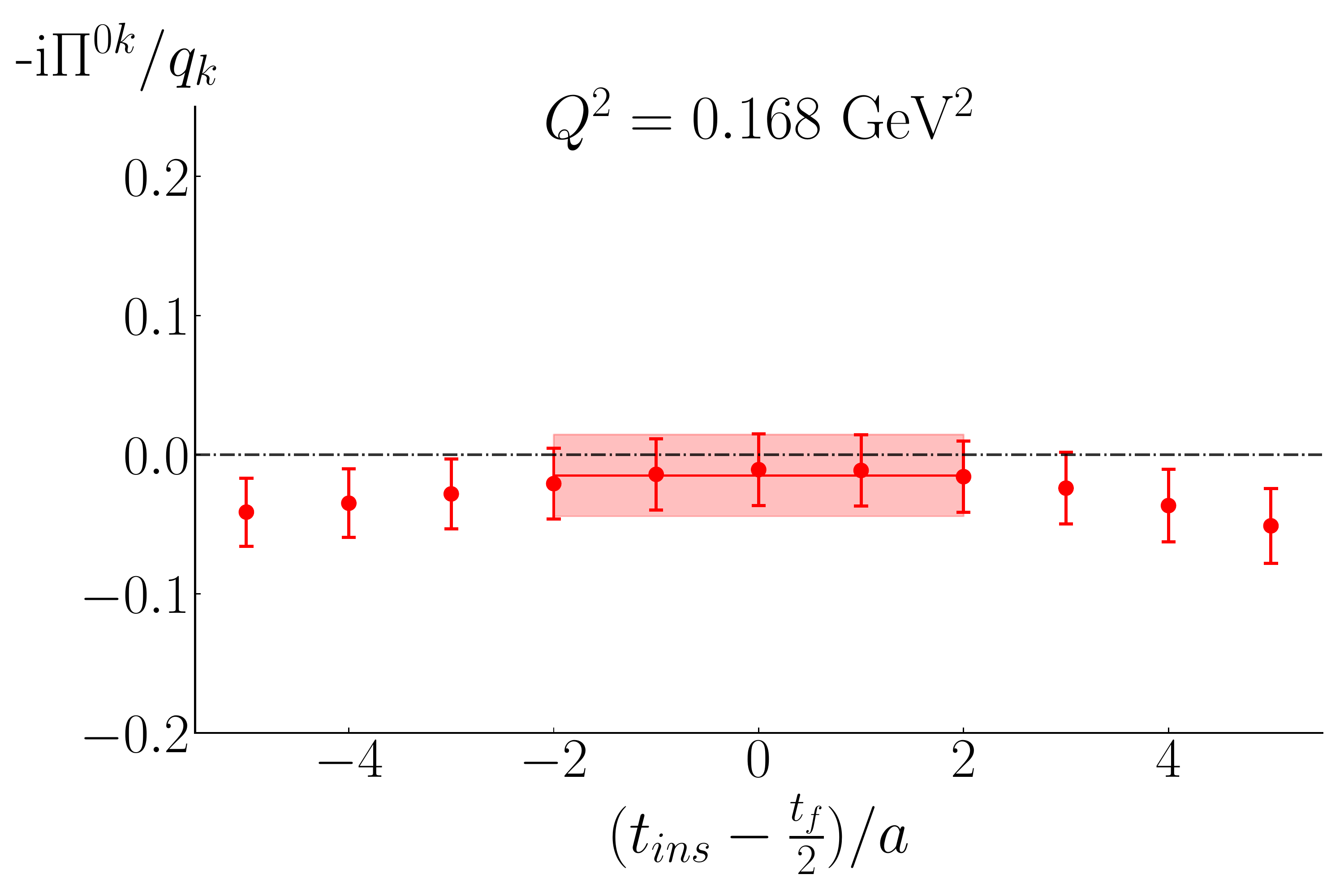}    
\end{minipage}
\captionof{figure}{Ratio of Eq.~\eqref{eq:Pi_munu} as a function of insertion time \(t_{ins}\) at fixed sink-source time separation \(t_f=12a\). The three smallest values of the  momentum transfer squared are shown. In the first row, we show results using  the gluonic  definition of \(\mathcal{Q}\) and  (\(\tau_{\rm flow}=3.5\)), while in the second row, the results are obtained using the spectral projectors for the computation of \(\mathcal{Q}\) and (\(M_{\rm thr}=64.98\) MeV). The bands are  the result of a constant fit in the plateau region excluding symmetrically 3 and 4 time slices for the gluonic (top panel) and fermionic (bottom panel) definition of \(\mathcal{Q}\), respectively.}
\label{fig:tins_plateau}
\end{figure}

From Eq.~\eqref{eq:G3ptQ_prjcted} and Eq.~\eqref{eq:G3ptQ_prjcted} we define
\be
\Pi^{\mu k}_{3pt,\mathcal{Q}}(\vec{q}) \equiv \lim_{t_f,t_{ins}\rightarrow\infty} \frac{G^{\mu}_{3pt,\mathcal{Q}}(\Gamma_k,\vec{q},t_f,t_{ins}) }{ G_{2pt}(\Gamma_0,\vec{0},t_f) } R_{2pt}\,, \label{eq:Pi_munu}
\ee
where \(\vec{p}_f=\vec{0}\) and $R_{2pt}$ reads  
\be
R_{2pt} \equiv \sqrt{\frac{G_{2pt}(\Gamma_0,\vec{q},t_f-t_{ins}) G_{2pt}(\Gamma_0,\vec{0},t_{ins}) G_{2pt}(\Gamma_0,\vec{0},t_f) }{G_{2pt}(\Gamma_0,\vec{0},t_f-t_{ins})G_{2pt}(\Gamma_0,\vec{q},t_{ins}) G_{2pt}(\Gamma_0,\vec{q},t_f)}}\,.
\label{eq:Rratio}
\ee
This ratio cancels unknown overlaps and exponential time dependence at large $t_f$ and $t_{\rm ins}$ times. 
Following the same steps that lead to Eq.~(55) of Ref.~\cite{Alexandrou:2015spa} with correction proposed in Ref.~\cite{Abramczyk:2017oxr}, one can express $\Pi^{0k}_{3pt,\mathcal{Q}}$ in terms of form factors as follows
\begin{equation}
\Pi^{0k}_{3pt,\mathcal{Q}}(\vec{q}) = \frac{i q_k \mathcal{C}}{2m_N} \left(\alpha_N  G_E(Q^2)-\frac{F_3(Q^2)}{2m_N} (E_N+m_N)   \right)\,,
    \label{eq:F3_extract}
\end{equation}
where \(E_N\) is the initial energy of the nucleon, \(\mathcal{C}=\sqrt{(2m_N^2)/(E_N(E_N+m_N))}\) is a kinematic factor and \(G_E(Q^2) = F_1(Q^2) + (q^2/(2m_N^2)) F_2(Q^2)\) is the electric Sachs form factor. It can be obtained from \(\Pi^{00}_{3pt}\) (see  Eq.~(A4) of Ref.~\cite{Alexandrou:2018sjm}).
We invert Eq.~\eqref{eq:F3_extract} to extract $F_3$ form factor  $\Pi^{00}_{3pt}$ and $\Pi^{0k}_{3pt,\mathcal{Q}}$ ratios. While the former exhibits a clear signal~\cite{Alexandrou:2018sjm}, the latter is affected by large uncertainties. This can be seen in Fig.~\ref{fig:tins_plateau}, where values of \(\Pi^{0k}_{3pt,\mathcal{Q}}\) as a function of the insertion time \(t_{ins}\) are reported for the smallest three non-zero values of the momentum transfer squared, i.e. \(Q^2 = 0.056\ \text{GeV}^2\), \(Q^2 = 0.111\ \text{GeV}^2\) and \(Q^2 = 0.164\ \text{GeV}^2\)
Data reported are at fixed sink-source time separation \(t_f=12a\) and we averaged among momenta with non-zero \(k\)-component in all \(k\)-directions. We fit the ratio in symmetric intervals \([-t_{\rm fit},t_{\rm fit}]\) and vary the fit ranges, taking  \(t_{\rm fit}=2,3,4\). However, the systematic calculated in this way is negligible if compared to the large statistical uncertainty.

\begin{figure}[!h]
\begin{minipage}[c]{0.49\linewidth}
\includegraphics[width=0.95\textwidth]{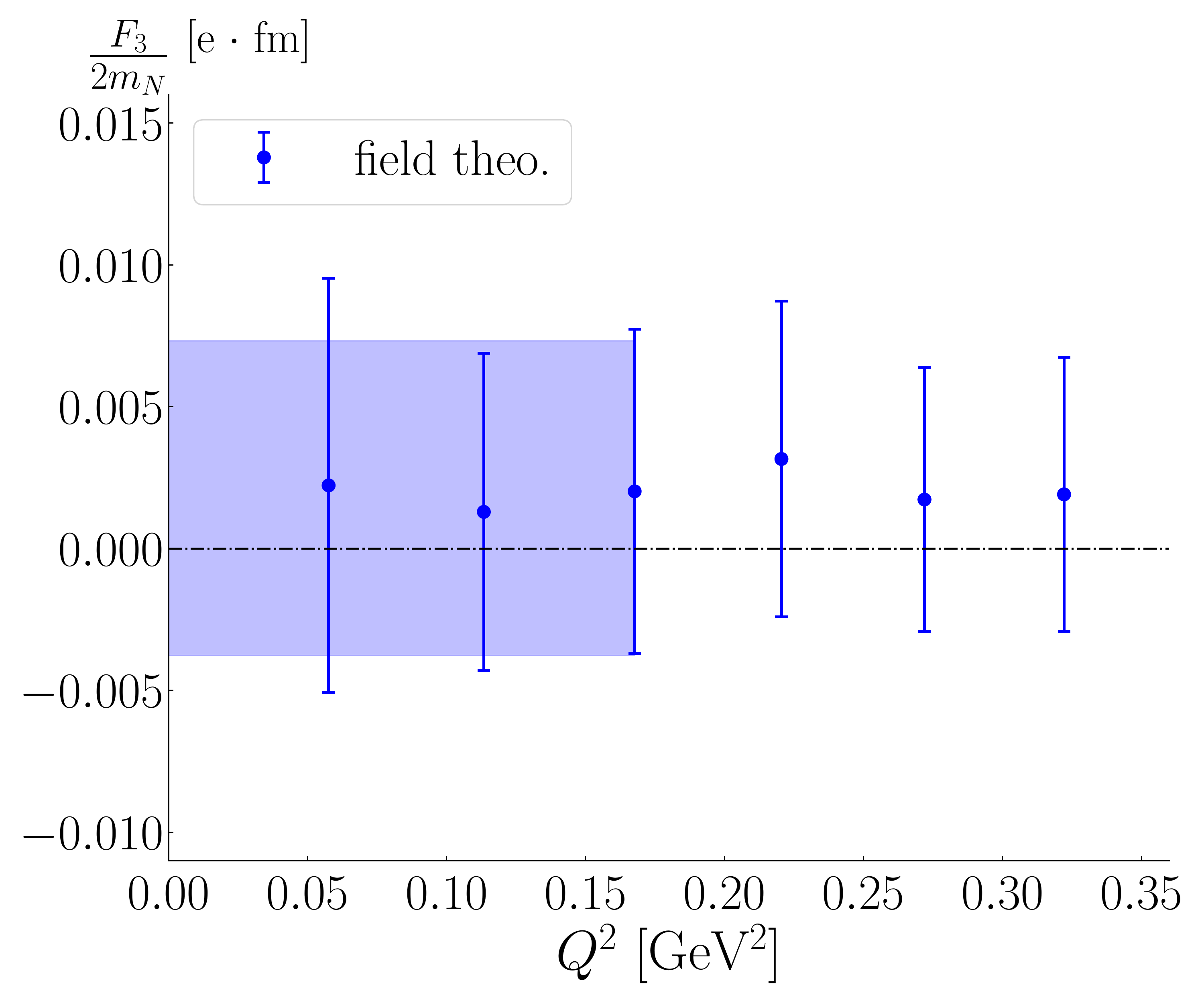}
\end{minipage}%
\hfill
\begin{minipage}[c]{0.49\linewidth}
\includegraphics[width=0.95\textwidth]{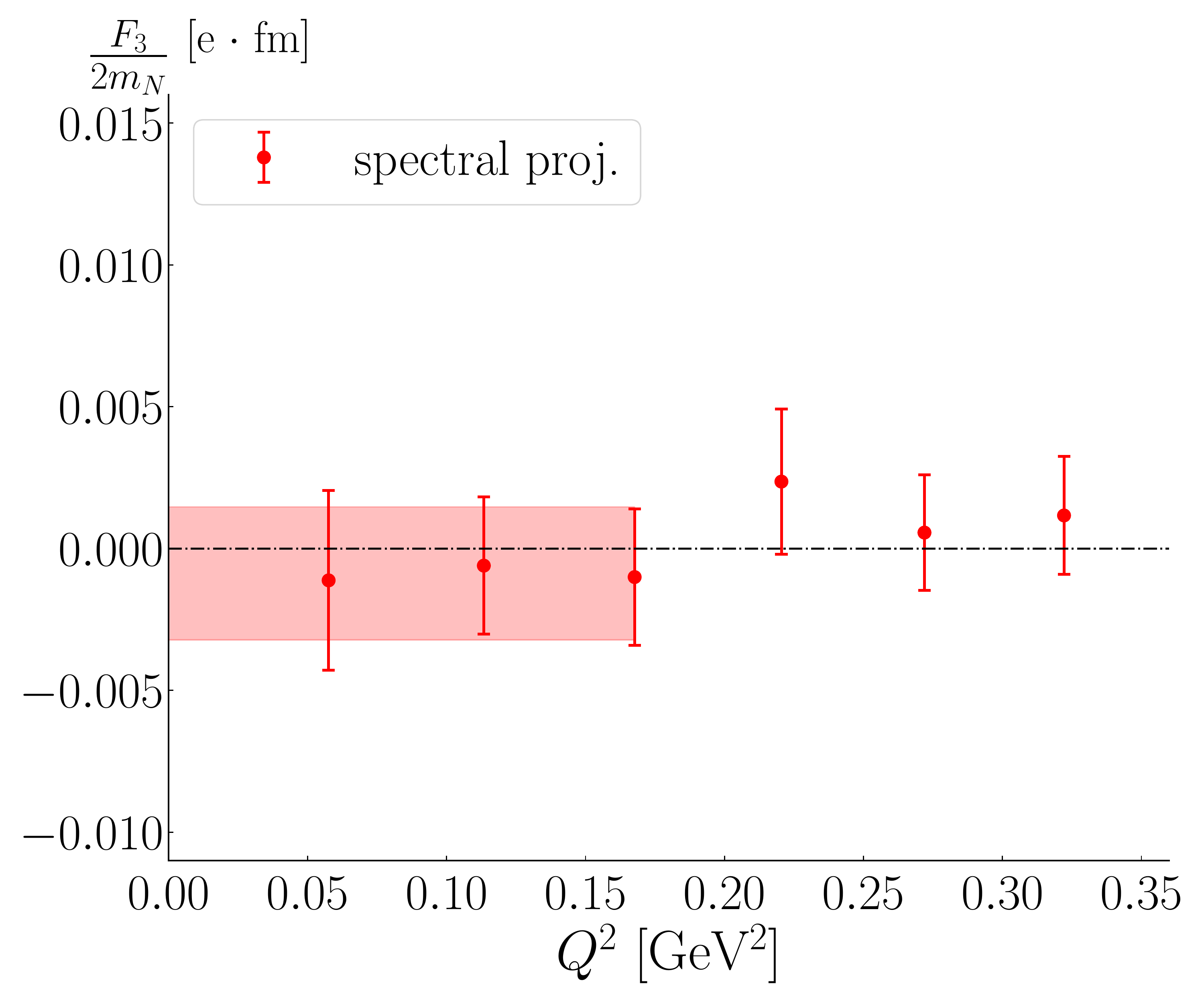}
\end{minipage}
\captionof{figure}{\(F_3(Q^2)\) as a function of \(Q^2\), using a the field theoretical or gluonic definition of the topological charge (left panel) and the fermionic definition based on spectral projectors (right panel). The blue and red   bands represent the weighted average of the values at the 3 smallest \(Q^2\) values.  }
\label{fig:F3_q0}
\end{figure}

In Fig.~\ref{fig:F3_q0} we report $F_3(Q^2)$ as a function of \(Q^2\). We take the weighted average of the values at the three smallest \(Q^2\) as the extrapolation of $F_3(Q^2)$ to $Q^2=0$. More involved fit forms are not viable with this level of uncertainty.

This leads to our final results for $d_N^{\theta}$:
\begin{align}\label{eq:result_g}   
\text{field~theoretical~or~gluonic definition} \qquad \ \lvert d_N^{\theta}\rvert = 0.0018(56) \: \theta \: {\rm e}\cdot{\rm fm} \,,\\
\text{fermionic definition via spectral~projectors} \qquad \ \lvert d_N^{\theta}\rvert = 0.0009(24)\: \theta \: {\rm e}\cdot{\rm fm}\,.
\label{eq:result_f}   
\end{align}

\begin{figure}[!h]
\begin{minipage}[c]{0.33\linewidth}
\centering
\includegraphics[width=1.\textwidth]{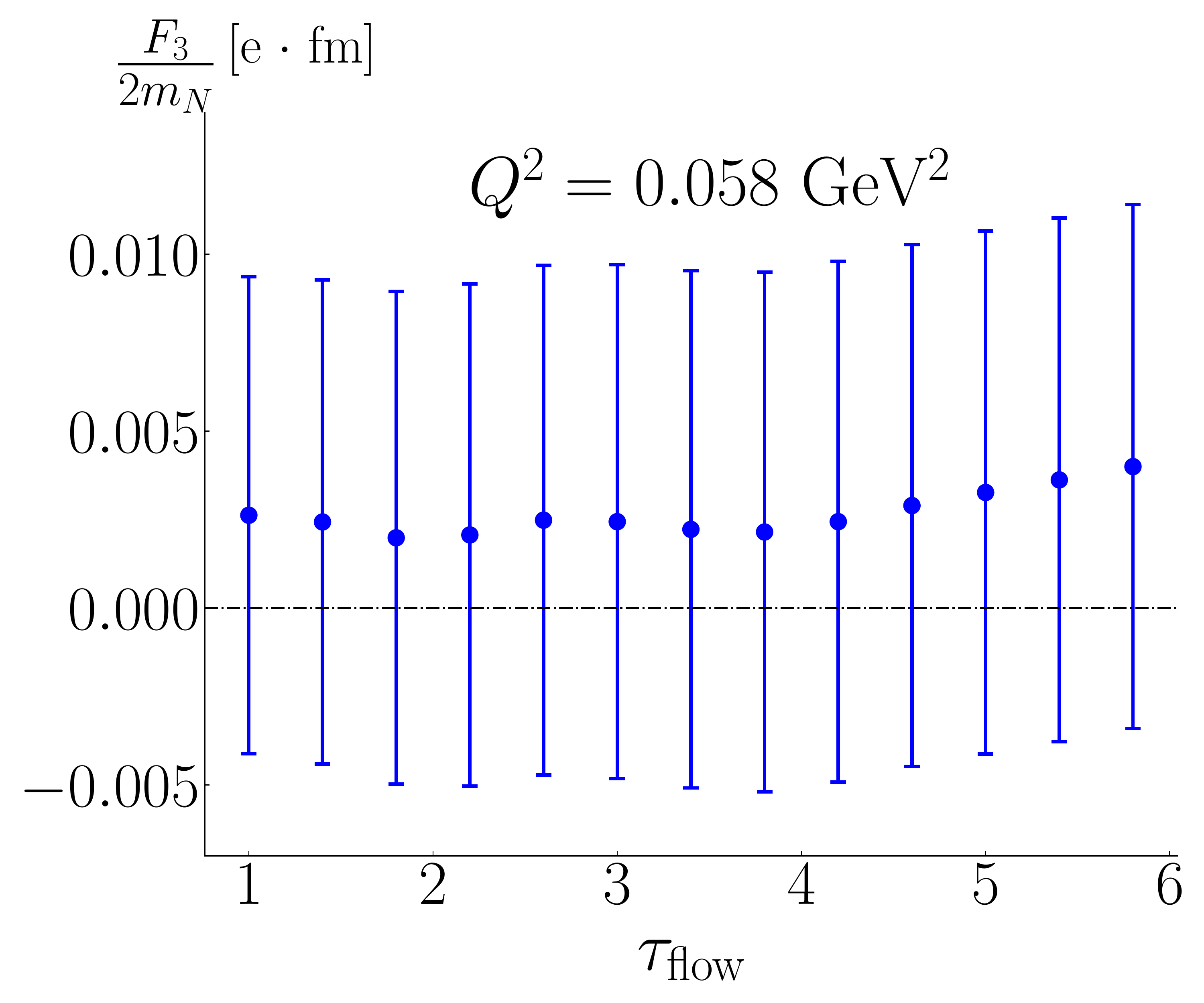}
\end{minipage}%
\hfill
\begin{minipage}[c]{0.33\linewidth}
\centering
\includegraphics[width=1\textwidth]{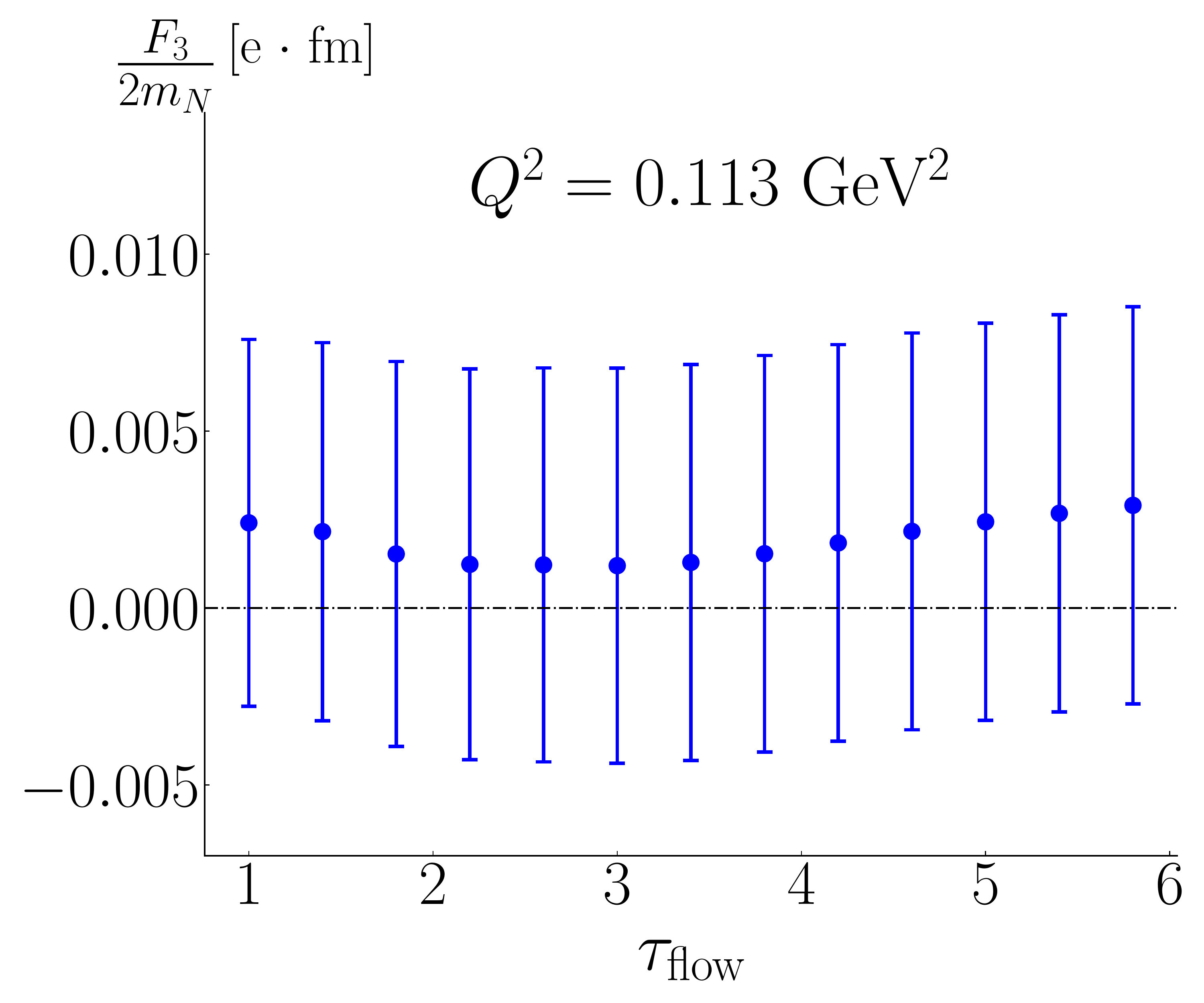}
\end{minipage}%
\hfill
\begin{minipage}[c]{0.33\linewidth}
\centering
\includegraphics[width=1\textwidth]{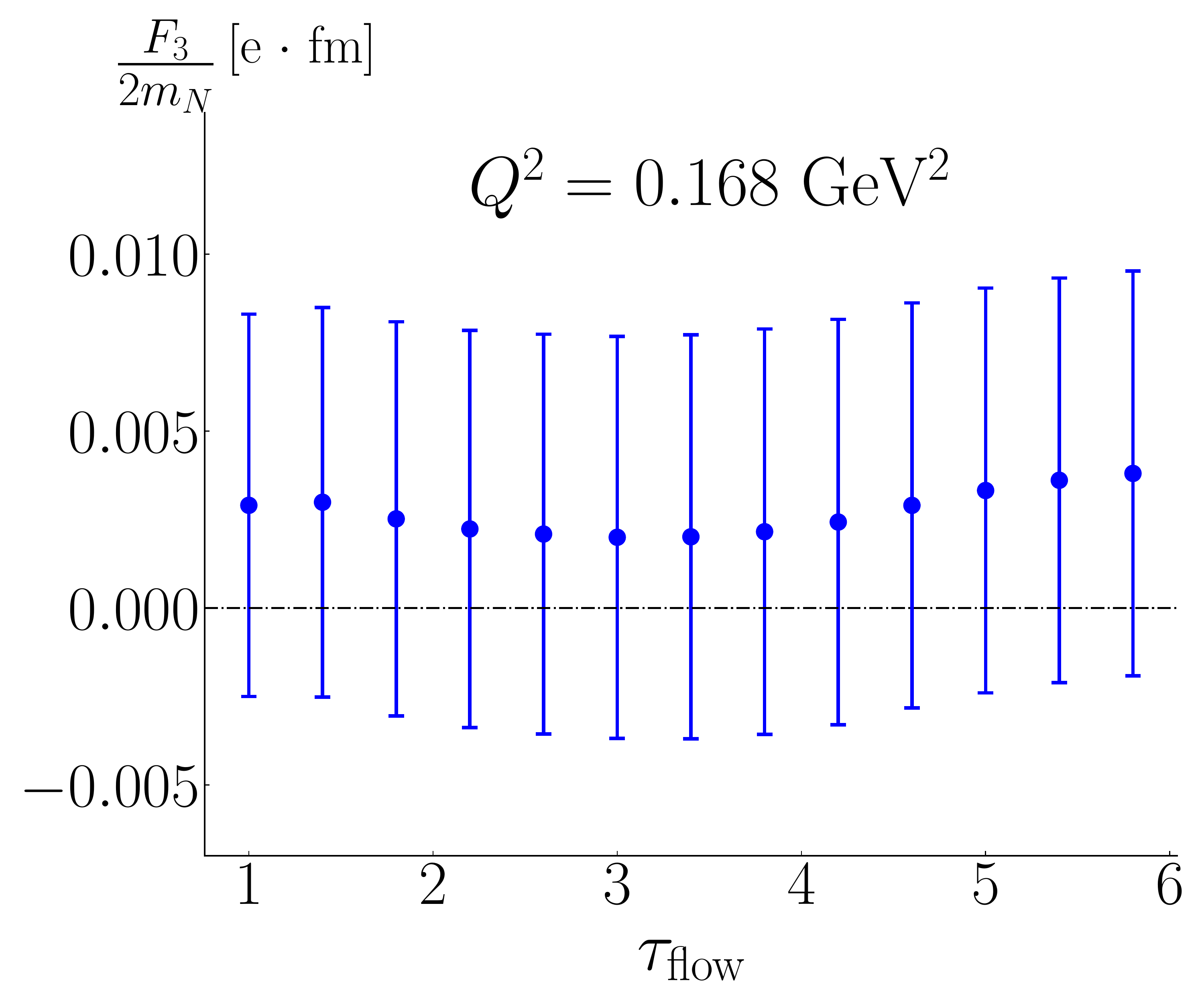}
\end{minipage}
\begin{minipage}[c]{0.33\linewidth}
\centering
\includegraphics[width=1.\textwidth]{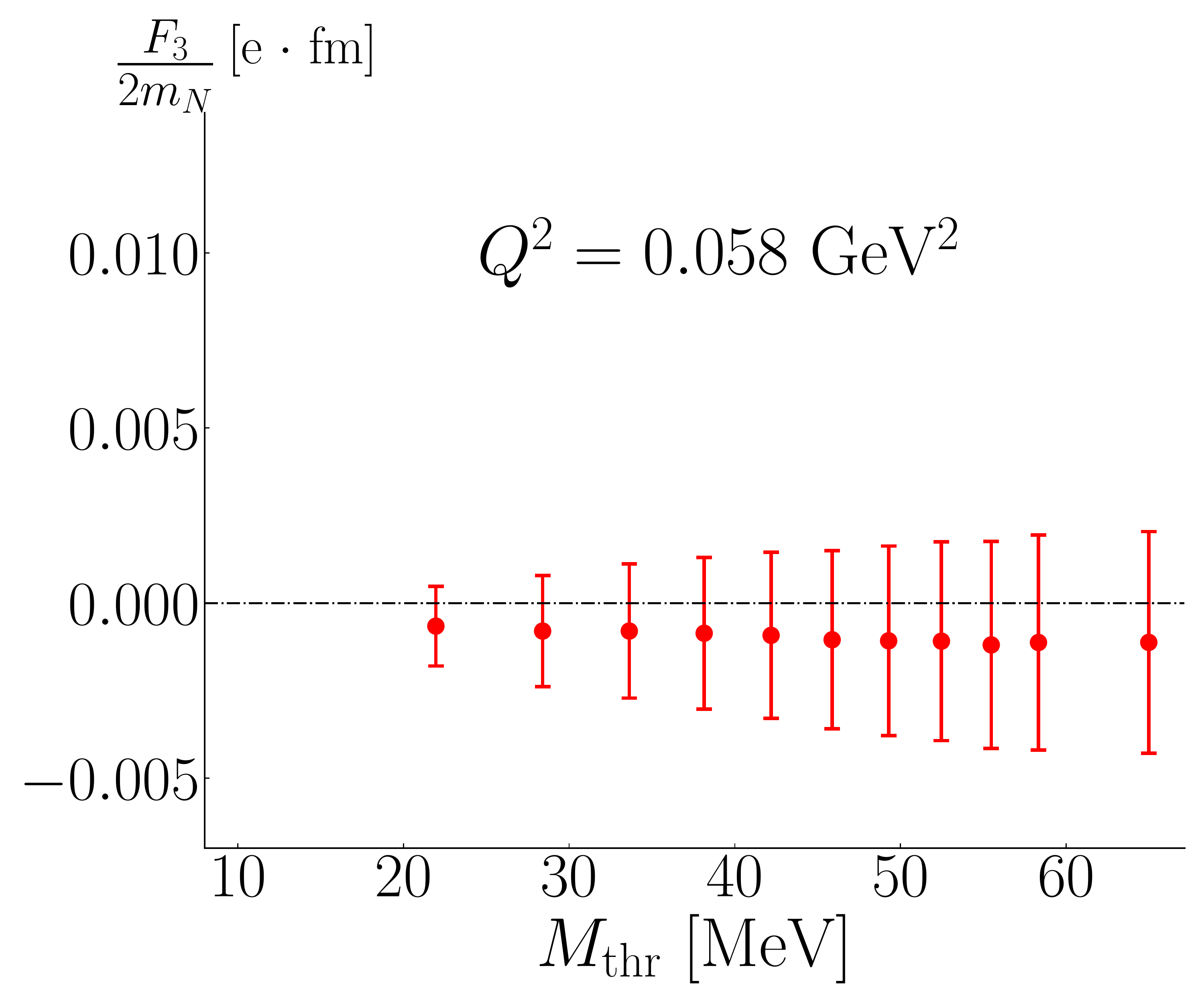}
\end{minipage}%
\hfill
\begin{minipage}[c]{0.33\linewidth}
\centering
\includegraphics[width=1.\textwidth]{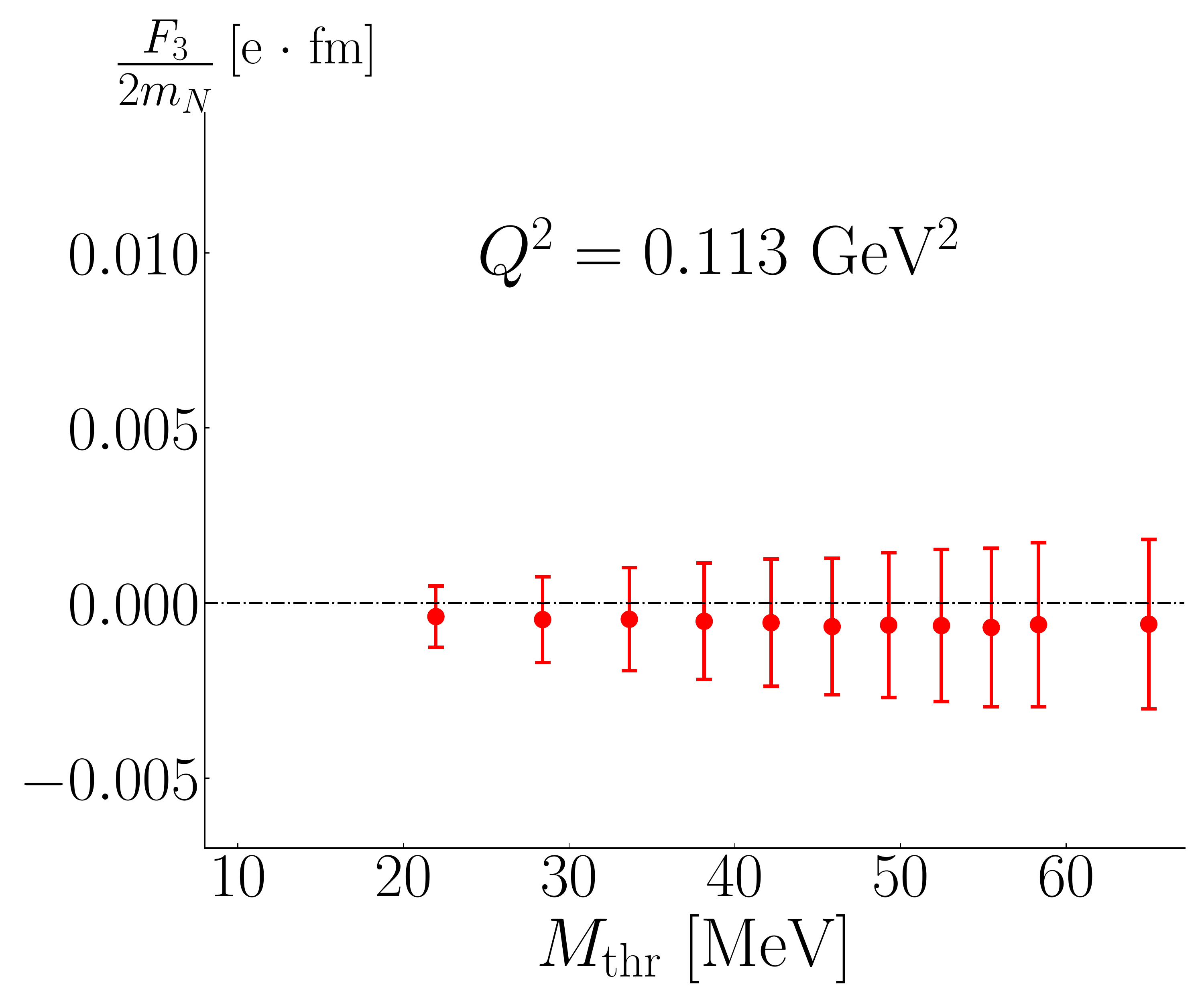}
\end{minipage}%
\hfill
\begin{minipage}[c]{0.33\linewidth}
\centering
\includegraphics[width=1.\textwidth]{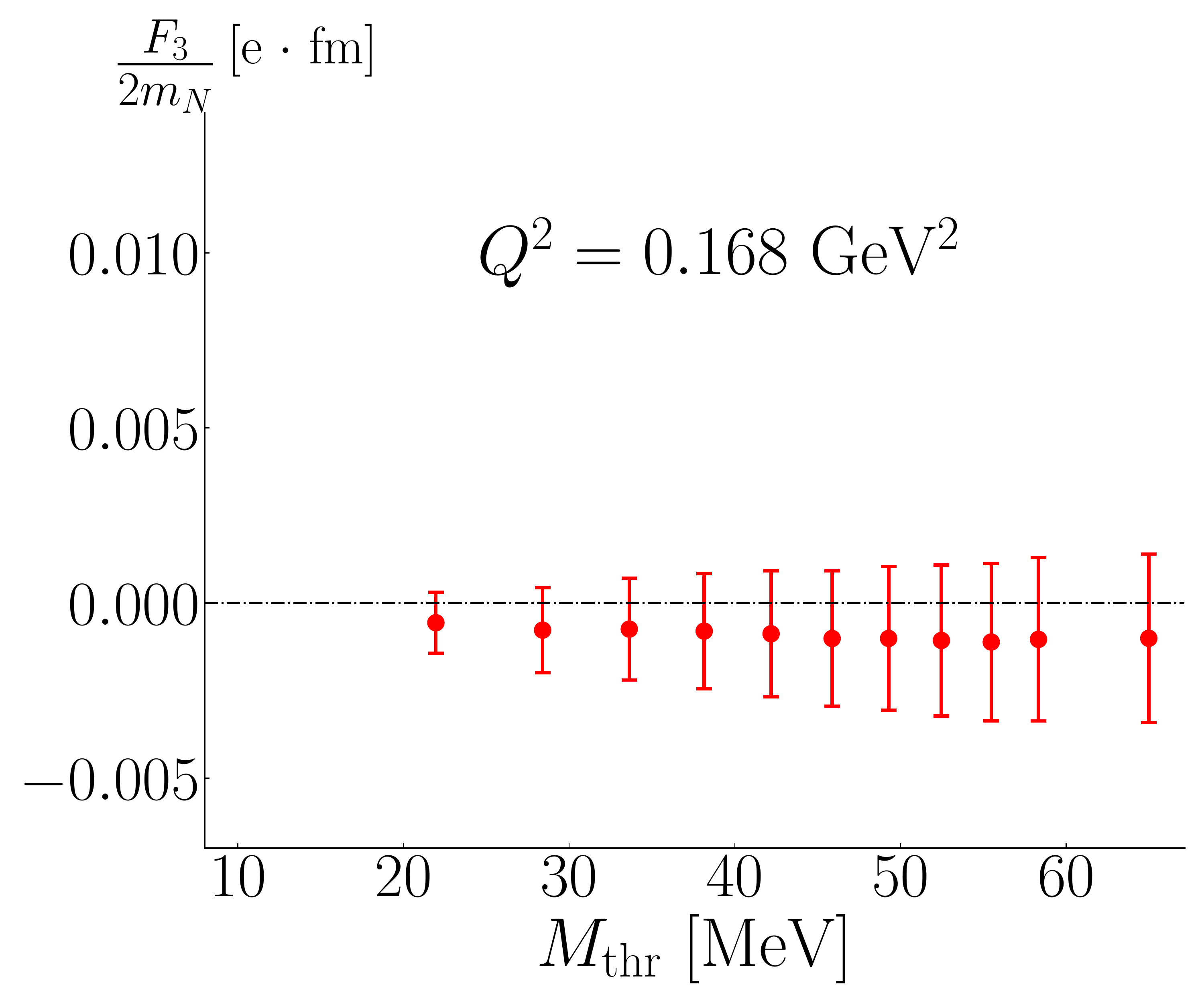}
\end{minipage}
\captionof{figure}{ Dependence of the \(F_3(0)\) on the smoothing scale $\tau_{\rm flow}$ for the gluonic (upper row) and cut-off \(M_{\rm thr}\) for the fermionic (bottom row) definitions used in computation of the topological charge, for the three smaller values of the momentum transfer squared.}
\label{fig:F3_d}
\end{figure}

If we take the absolute error as a bound for the magnitude of the nEDM,  we find that the definition of the topological charge via spectral projectors is  \(2\) times more accurate than that from  the gluonic definition. Therefore, the additional cost due to the computation of the eigenmodes for the fermionic definition of \(\mathcal{Q}\), is compensated by the increased precision.
The dependence of $F_3(Q^2)$ using spectral projectors on the cut-off \(M_{\rm thr}\) shows that  the mean value of the form factor does not depend on \(M_{\rm thr}\) and only the error increases with increasing \(M_{\rm thr}\), as shown in Fig.~\ref{fig:F3_d}.

\section{Conclusions}
\label{sec:comparison}
\begin{figure}[!h]
    \centering
    \includegraphics[width=\textwidth]{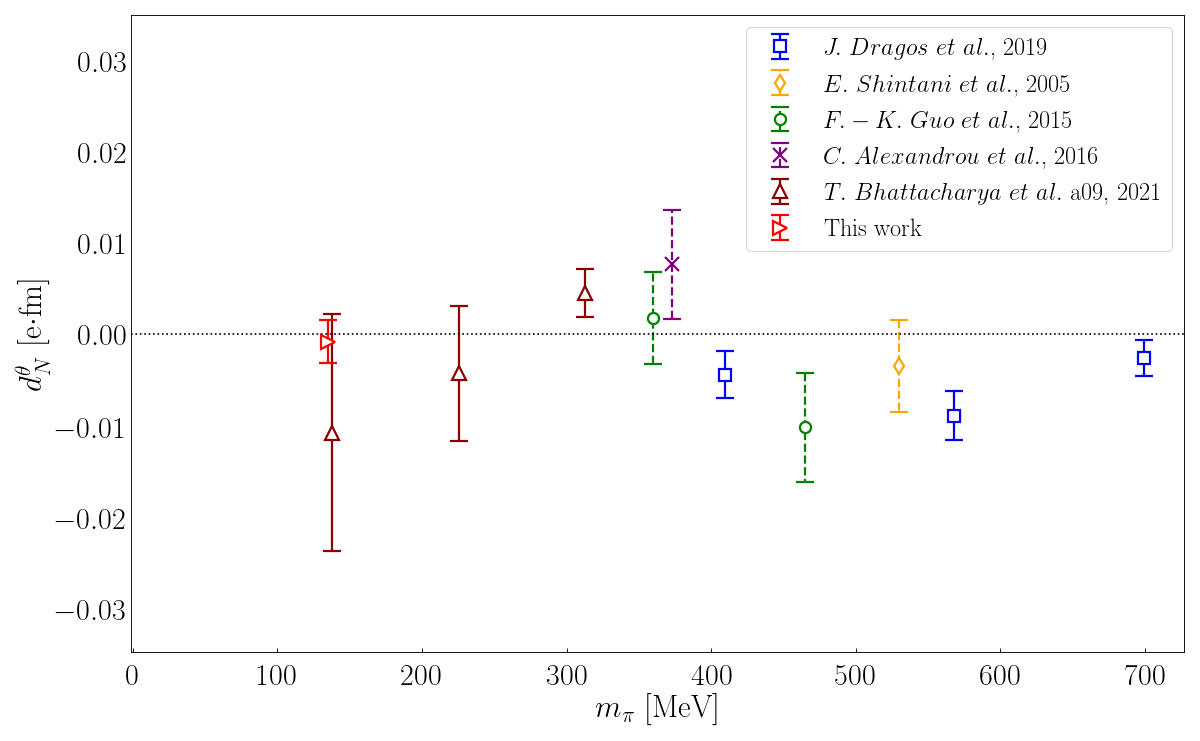}
    \captionof{figure}{Comparison with other lattice QCD determinations of nEDM. Values from Refs.~\cite{Shintani:2005xg,Guo:2015tla,Alexandrou:2015spa} (dashed error bars)  are corrected using  Table~III of Ref.~\cite{Abramczyk:2017oxr}, where the spurious contribution coming from \(F_2(Q^2)\) is subtracted. See Ref.~\cite{Abramczyk:2017oxr} for further details.}\vspace*{-0.3cm}
    \label{fig:comparison}
\end{figure}

 A comparison of our result with those of other recent lattice QCD studies is shown in  Fig.~\ref{fig:comparison} we provide a for a similar lattice spacing. Our value is compatible with the result at the physical point presented in Ref~\cite{Bhattacharya:2021lol}, that, however, shows a 4 times larger uncertainty. 
 Since the errors grow with decreasing pion mass and so does the computational cost, achieving such an accuracy it is a major outcome of our work. Moreover, the accuracy of the extrapolation of the nEDM to the physical point obtained in Ref.~\cite{Dragos:2019oxn}, i.e. \(\lvert d_N^\theta \rvert = 0.00152(71) \theta\:{\rm e}\cdot{\rm fm}\), is due to the chiral continuum extrapolation, where systematic errors from using chiral expressions cannot be determined. Their actual data have uncertainties similar to this work.

We found that the fermionic definition of $\cal Q$ leads to a two-fold increase in statistical accuracy in the determination of the nEDM as compared to the gluonic  definition. This allows us to obtain a value for the nEDM at the physical point to unmatched precision 
   \begin{eqnarray}
    \ \lvert d_N^{\theta}\rvert = 0.0009(24) \; \theta \; e \, {\rm fm}\,.
    \end{eqnarray}
 Ruling out a zero value would require at least a 2-orders-of-magnitude increase in statistics.  Alternative approaches, like using configurations generated with an imaginary $\theta$-term will be considered in future  investigations.

\section{Acknowledgements}
A.A. is supported by the European Union's Horizon 2020  programme ``Tips in SCQFT'' under the Marie Sk\lpol odowska-Curie (MAC) grant agreement No. 791122.   K.H. is financially supported by the Cyprus Research  and Innovation  foundation  under  contract  number  POST-DOC/0718/0100. A.T. is a MSC fellow funded by the European Union’s Horizon 2020 programme under grant agreement No 765048. Results were obtained using Piz Daint at Centro Svizzero di Calcolo Scientifico (CSCS), via the project with id s702, Marconi100 at CINECA, Italy,  within the PRACE project with Id Pra20\_5171 and   the Juwels system at the research center in J\"{u}lich, under the project with id CHCH02.

\vspace*{-0.3cm}
\bibliographystyle{JHEP}
\bibliography{biblio1}
\end{document}